\documentclass[12pt]{article}
\usepackage{amssymb}
\def\lddots{\mathinner{\mkern1mu\raise1pt\hbox{.}\mkern2mu
\raise4pt\hbox{.}\mkern2mu\raise7pt\vbox{\kern7pt\hbox{.}}\mkern1mu}}
\makeatletter
\def\numberbysection{\@addtoreset{equation}{section}
\def\theequation{\thesection.\arabic{equation}}}
\makeatother

\usepackage{epic}
\usepackage{eepic}
\usepackage{graphicx}
\usepackage{latexsym}
\usepackage{amssymb}
\numberbysection

\newcommand{\be}{\begin{eqnarray}}
\newcommand{\ee}{\end{eqnarray}}
\newcommand{\non}{\nonumber}

\begin{document}

\begin{titlepage}
\strut\hfill
%\vspace{.5in}
%
\begin{center}
{\bf {\Large Non-diagonal reflection for the non-critical XXZ model}}
\\[0.5in]

\vskip 0.2cm

{\large Anastasia Doikou\footnote{e-mail: doikou@bo.infn.it}}

\vspace{10mm}

{\small University of Bologna, Physics Department, INFN Section\\
Via Irnerio 46, 40126 Bologna, Italy}

\end{center}
%
%\vspace{.5in}

\vfill

\begin{abstract}

The most general physical boundary $S$-matrix for the open XXZ spin chain
in the non-critical regime ($\cosh (\eta ) >1$) is derived starting from the
bare Bethe ansazt equations. The boundary $S$-matrix as expected is
expressed in terms of $\Gamma_q$-functions. In the isotropic limit corresponding
results for the open XXX chain are also reproduced.

\end{abstract}

\vfill

\baselineskip=16pt

\end{titlepage}

\section{Introduction}

The open XXZ model is considered as one of the prototype models in
describing a plethora of interesting boundary phenomena, and as such has
attracted much attention especially after the derivation of the
spectrum in the generic case where non-diagonal boundary magnetic
fields are applied \cite{chin}--\cite{galleas}. Our
objective in the present study is to derive from first principles
the most general physical boundary $S$-matrix for the XXZ chain in
the non-critical (massive) regime.

Diagonal boundary $S$-matrices for the open XXZ model in the
non-critical regime were extracted in \cite{jimbo} using
vertex-operator techniques, while parallel results obtained in
\cite{done1} from the Bethe ansatz point of view (see e.g. 
\cite{FT, korepin, AD}).
Similarly, diagonal reflection matrices
were derived in \cite{FS, donebr} for the critical XXZ model,
corresponding to the sine-Gordon boundary $S$-matrix for `fixed'
boundary conditions \cite{GZ}. After the derivation of the exact
spectrum and Bethe equations for the XXZ chain with non-diagonal
boundaries the generic boundary $S$-matrix for the critical XXZ
chain was computed in \cite{doikounew, raj}, corresponding to the
boundary $S$-matrix of sine-Gordon model \cite{GZ} for `free'
boundary conditions. A relevant discussion on the generic breather
boundary $S$-matrix within the XXZ framework may be also found in
\cite{doikounew}. Note also that analogous results regarding diagonal and
non-diagonal solitonic boundary $S$-matrices were formulated in \cite{ddvb, nepoh}
using the so called nonlinear integral equation (NLIE) method
\cite{ddv}.

To extract the generic boundary $S$-matrix for the non-critical XXZ model we follow 
the logic of \cite{doikounew}, i.e. we focus on
the open chain with a trivial left
boundary and a generic non-diagonal
right boundary associated to the full $K$-matrix \cite{GZ, DVGR} .
As also noted in \cite{doikounew} the main advantage of the approach adopted
--considering special boundary conditions--
is that one eventually
deals with a simple set of Bethe ansatz equations
similar to the ones of the XXZ chain with two diagonal boundaries.
Thus all relevant computations are drastically simplified (see also \cite{doikounew}), and one may follow 
the logic described in \cite{GMN, DMN, done1, donebr} for purely diagonal boundary
magnetic fields. Ultimately, the boundary $S$-matrix eigenvalues are extracted directly from the 
Bethe equations and are expressed in terms of $\Gamma_q$-functions ($q = e^{-\eta}$) \cite{GR} 
as in \cite{jimbo, done1}, where only diagonal boundaries are assumed. In the isotropic 
limit $q \to 1$ the corresponding  rational boundary $S$-matrix for the 
XXX open chain is also recovered \cite{macsh}.

\section{Bethe ansatz and boundary $S$-matrix}

Before we proceed with the Bethe ansatz analysis it will be useful
for our purposes here to give the explicit expressions of the
right and left boundary $K$-matrices that give rise to the open
Hamiltonian under consideration: \be {\cal H} &=& -{1\over
4} \sum_{i=1}^{N-1}\Big (\sigma_{i}^{x}\sigma_{i+1}^{x}
+\sigma_{i}^{y}\sigma_{i+1}^{y}+ \cosh(\eta)\
\sigma_{i}^{z}\sigma_{i+1}^{z}\Big ) -{N\over 4}\ \cosh (\eta)
+{\sinh (\eta) \over 4}\sigma_{N}^{z} \non\\
& +& {\sinh (\eta) \cosh (\eta \xi) \over 4 \sinh(\eta \xi)}
\sigma_{1}^{z} - {\kappa \sinh (\eta) \over 2 \sinh (\eta
\xi)}\Big (\cosh( \eta \theta) \sigma_{1}^{x} +i\sinh ( \eta
\theta)\sigma_{1}^y \Big ) \label{H0} \ee where in the
non-critical regime we are focusing here $~\cosh(\eta) > 1$, also
$\sigma^{x,y,z}$ are the $2 \times 2$ Pauli matrices, and the
boundary parameters $\xi,\ \kappa,\ \theta$, are the free
parameters of the generic $K$-matrix \cite{GZ, DVGR}, which will
be introduced subsequently. 

To obtain such a Hamiltonian we consider the open chain
constructed using Sklyanin's formalism \cite{sklyanin}, with left
boundary $K^+ \propto {\mathbb I}$ and right boundary associated
to the general solution of the reflection equation
\cite{cherednik} given in \cite{GZ, DVGR} i.e. \be
K^{-}(\lambda )= \left( \begin{array}{cc}
\sin [\eta (-\lambda +i \xi)] e^{i\eta \lambda} & \kappa e^{\eta \theta } \sin (2\eta \lambda) \\
\kappa e^{-\eta \theta } \sin (2\eta \lambda) & \sin [\eta (\lambda +i \xi)] e^{-i\eta \lambda}   \\
\end{array} \right). \label{def} \ee The latter $K$-matrix has two eigenvalues given below:
\be && \varepsilon_1(\lambda) = 2  \kappa\ \sin[\eta (\lambda +
ip^+)]\ \sin [\eta (\lambda +ip^-)] \non\\ &&
\varepsilon_2(\lambda) = 2 \kappa\ \sin[\eta (\lambda - ip^+)]\
\sin [\eta (\lambda -ip^-)] \label{eigenv} \ee where the
parameters $p^{\pm}$ are defined as: \be {e^{\pm \eta \xi} \over
2\kappa} = i \cosh [\eta (p^+ \pm  p^-)]. \label{param} \ee
Note that we assume here the parametrization 
used in \cite{GZ} in the sine-Gordon context, 
(see also \cite{doikounew} the references therein).
Such a parametrization is also quite practical within the
Temberley-Lieb algebra framework \cite{doma}.
The parameter $\theta$ appearing in (\ref{def}) may be removed 
by means of a simple gauge transformation, that leaves the XXZ $R$-matrix invariant, 
and henceforth we consider it for
simplicity to be zero (see also \cite{GZ, doikounew}). The
$K$-matrix (\ref{def}) may be easily diagonalized by virtue of a
constant ($\lambda$-independent) gauge transformation: \be
\mbox{diag} \Big (\varepsilon_1(\lambda),\ \varepsilon_2(\lambda)\Big ) = {\cal
M}^{-1}(p^+,\ p^-)\ K(\lambda)\ {\cal M}(p^+,\ p^-) \label{gauge}
\ee where ${\cal M}$ is defined as: \be {\cal M}(p^+,\ p^-) =
\left( \begin{array}{cc}
 1                         & 1 \\
 i e^{ \eta (p^+ + p^-)}  & i e^{- \eta (p^+ + p^-)}  \\
\end{array} \right). \ee Note that the above transformation modifies 
dramatically the XXZ $R$-matrix, so it is not possible to simply implement a global gauge 
tranformation changing the basis in order to diagonalize the open transfer matrix as in e.g. \cite{annecy2}. 
It is also worth pointing out the
similarity between the matrix ${\cal M}(p^+,\ p^-)$ and the local
gauge transformation employed for the diagonalization of the open
XXZ transfer matrix with non-diagonal boundaries \cite{chin, doikouj, doikous}.

We recall now the exact Bethe ansatz equations for the
open XXZ chain in the case of a right non-diagonal boundary and a left trivial
diagonal. The Bethe equations in this case reduce to the following
simple form (see also \cite{doikounew}):
\be
&& {\sin  [\eta ( \lambda_i - {i \over 2}(2p^+ +1) )]
\over
\sin  [\eta \left( \lambda_i + {i\over 2} (2p^++1)\right)] }\ {\sin 
[\eta ( \lambda_i - {i \over 2}(2p^- +1) )]
\over
\sin  [\eta \left( \lambda_i + {i\over 2} (2p^-+1)\right)] }\
\ {\cos  [\eta \left( \lambda_i + {i\over 2} \right)]
\over
\cos  [\eta \left( \lambda_i - {i\over 2} \right)] } 
\Bigg( {\sin  [\eta ( \lambda_i + {i \over 2})]
\over
\sin  [\eta \left( \lambda_i -{i\over 2} \right )] }\Bigg )^{2N+1}
\non\\ && = - \prod_{j = 1}^{M} {\sin  [\eta ( \lambda_i -\lambda_j + i )]
\over
\sin  [\eta ( \lambda_i-\lambda_j -i)] }\  {\sin  [\eta ( \lambda_i +\lambda_j + i )]
\over
\sin  [\eta ( \lambda_i+\lambda_j -i)] }. \label{BA}
\ee We consider here, without loss of generality $\eta > 0$,
$~p^{\pm} > {1 \over 2} $ and $Re \left( \lambda_{\alpha} \right)
\in [ 0 \,, {\pi\over 2 \eta}]\, ~\lambda_{\alpha} \ne 0,
{\pi\over 2 \eta}$  (see e.g. \cite{GMN} for details on this
restriction). For relevant results on various representations of
$U_q(sl_2)$ see \cite{doikouj, doikous, annecy}.

As pointed out in \cite{doikounew} the integer $M$ is associated
to a non-local conserved quantity ${\cal S}$, which has the same
spectrum as $S^z$ (for more details we refer the interested reader
to \cite{doikounew, doikouj} and references therein) i.e. 
\be M ={N \over 2} -{\cal S}_{\varepsilon},\ee the subscript $\varepsilon$ 
stands for the eigenvalue.

Our objective now is to explicitly derive the physical boundary
$S$-matrix, and in particular the relevant overall physical
factor, which provides in general significant information on the
existence of boundary bound states. We define the boundary
$S$-matrices ${\mathrm K}^{\pm}$ by the quantization condition
\cite{AD, GMN} \be \left( e^{i 2 p(\tilde\lambda) N} {\mathrm
K}^{+}\ {\mathrm K}^{-}- 1 \right) |\tilde \lambda \rangle = 0.
\label{quantizationopen} \ee $\tilde \lambda $ is the rapidity of
the `hole' --particle-like excitation, and $p(\tilde \lambda)$ is
the momentum of the hole. 

The density of a state is obtained in a standard way from the Bethe ansatz equations 
after taking the  $\log$ and the derivative \cite{korepin, AD, GMN, DMN, done1}. 
More precisely the Fourier transform of the density for the one-hole state
turns out to be: \be \hat \sigma_{s}(\omega) &=& 2 \hat  \epsilon(\omega) +
{1\over N} {\hat a_{2}(\omega)\over 1 + \hat
a_{2}(\omega)}
(e^{i\omega \tilde \lambda}+ e^{-i\omega \tilde \lambda}) \non\\
&+& {1\over N}{1 \over 1 + \hat a_{2}(\omega)} \bigg[ \hat
a_{1}(\omega) + \hat a_{2}(\omega) + \hat
b_{1}(\omega) - \hat a_{2p^- +1}(\omega)
- \hat a_{2p^+ + 1}(\omega) \bigg], \non\\
\label{dens} \ee where we define the following Fourier transforms
\be \hat a_{n}(\omega) = e^{-\eta n |\omega|}, \qquad
\hat b_{n}(\omega) = (-)^{\omega}\ \hat a_{n}(\omega), \qquad
\hat \epsilon(\omega) = {\hat a_1(\omega) \over 1 + \hat a_2(\omega) }=
{1 \over 2 \cosh({\omega \over 2})} \label{fourier3} \ee
$\epsilon(\tilde \lambda)$ corresponds also to the energy of the particle-like excitation.
The similarity of the latter formula
(\ref{dens}) with the one obtained in
the case of two diagonal boundaries \cite{GMN, done1} is indeed noticeable.
This is a crucial point enabling a
simplified derivation of the boundary $S$-matrix. In our case both terms
depending on $p^{\pm}$ are assigned to the right boundary,
otherwise one follows the logic of the fully diagonal case (see e.g.
\cite{GMN, done1}).

The boundary matrix ${\mathrm K}^-$, of the generic form
(\ref{def}), has two eigenvalues ${\mathrm k}_{1,2}$, whereas the left
boundary matrix is trivial ${\mathrm K}^+(\tilde \lambda) =
k_0(\tilde \lambda) {\mathbb I}$. From the density (\ref{dens}), 
the quantization condition (\ref{quantizationopen}),
and recalling that $\epsilon(\lambda) = {1\over 2 \pi} {d p(\lambda) \over  d \lambda}$ we can
explicitly derive the quantities $k_0,\ {\mathrm k}_{1,2}$ (see e.g. \cite{GMN, DMN, done1} for more details).  
Actually, the eigenvalues ${\mathrm k}_1(\tilde\lambda,\  p^\pm)$ and ${\mathrm k}_2(\tilde\lambda,\ p^{\pm})$
may be seen as the boundary scattering amplitudes for the one particle-like excitation
with ${\cal S} = +{1\over 2}$ and ${\cal S} = -{1\over 2}$,
respectively (see also \cite{doikounew} for more details)

We first compute the eigenvalue ${\mathrm k}_1$, which is expressed in terms of
the $\Gamma_{q}(x)$-function, --the $q$-analogue of the  Euler gamma function--
($q = e^{-\eta}$) defined \cite{GR} as
\be
\Gamma_{q}(x) = (1 - q)^{1-x} \prod_{j=0}^{\infty}
\left[ {\left( 1-q^{1+j} \right)\over \left(1-q^{x+j} \right)} \right]
\,, \qquad 0 < q < 1 \,.
\ee Using also the $q$-analogue of the
duplication formula \cite{GR}
\be
\Gamma_{q}(2 x)\ \Gamma_{q^{2}}({1\over 2}) =
(1 + q)^{2 x -1}\ \Gamma_{q^{2}}(x)\ \Gamma_{q^{2}}(x + {1\over 2}),
\ee
we obtain the following result for the first eigenvalue
${\mathrm k}_1(\tilde \lambda,\  p^+,\ p^-)$ (up to a constant phase factor):
\be {\mathrm k}_1(\tilde \lambda,\ p^+,\ p^-) &=& 2  \kappa\ \sin \Big
[\eta \Big (\tilde \lambda + {i\over 2}(2p^+ - 1)\Big )\Big ]\
\sin \Big [\eta \Big (\tilde \lambda + { i\over 2} (2p^- - 1)
\Big )\Big ] \non\\ &\times& k_0(\tilde \lambda)\ k_1(\tilde \lambda,\ p^+)\
k_1(\tilde \lambda,\ p^-)\ee where we define:
\be k_0(\tilde \lambda)=  q^{-4 i \tilde\lambda}
{\Gamma_{q^8} \left({-i\tilde\lambda\over 2} + {1\over 4}\right) \over
 \Gamma_{q^8} \left({i\tilde\lambda\over 2} + {1\over 4}\right)}\ {\Gamma_{q^8}
\left({i\tilde\lambda\over 2} + 1\right) \over \Gamma_{q^8}
\left({-i\tilde\lambda\over 2} + 1\right)} \ee \be
k_1(\tilde\lambda,\ x) = {(2\kappa)^{-{1\over 2}} \over \sin \Big
[\eta \Big (\tilde \lambda - {i\over 2}(2x - 1) \Big ) \Big ]}
{\Gamma_{q^4} \left({-i\tilde\lambda\over 2} + {1\over 4}(2 x
-1)\right) \over \Gamma_{q^4} \left({i\tilde\lambda\over 2} +
{1\over 4}(2 x-1)\right)} {\Gamma_{q^4} \left({i\tilde\lambda\over
2} + {1\over 4}(2 x +1)\right) \over \Gamma_{q^4}
\left({-i\tilde\lambda\over 2} + {1\over 4}(2 x +1)\right)}.
\non\\ \label{result1} \ee We turn now to the computation of the
second eigenvalue ${\mathrm k}_2(\tilde \lambda,\ p^{\pm})$,
corresponding to a one-hole state with ${\cal S} = -{1\over 2}$.
We implement the `duality' transformation \cite{sklyanin,
doikounew}, which modifies the boundary parameters $p^\pm
\rightarrow -p^\pm$ in the Bethe ansatz equations (\ref{BA}). This
transformation is the equivalent of deriving the Bethe ansatz
equations starting from the second reference state (the analogue
of the `spin down' state) \cite{annecy, refst}. Then we conclude
for the second eigenvalue: \be {{\mathrm k}_1(\tilde\lambda,\
p^+,\ p^-) \over {\mathrm k}_2(\tilde \lambda,\ p^+,\ p^-)} =
{\sin \Big [\eta \Big (\tilde \lambda + {i\over 2}(2p^+ - 1 )\Big
)\Big ]\ \sin \Big [\eta \Big (\tilde \lambda + {i\over 2}(2p^- -
1 )\Big )\Big ] \over \sin \Big [\eta \Big (\tilde \lambda -
{i\over 2}(2p^+ - 1 )\Big )\Big ]\ \sin \Big [\eta \Big (\tilde
\lambda -{i\over 2}(2p^- - 1 )\Big )\Big ]} \label{result2} \ee An
alternative way to extract the second eigenvalue is
instead of the `kink' state with ${\cal S} = -{1 \over 2}$, 
--after implementing the duality transformation-- to consider
the anti-kink state consisting of a hole and a
two-string state. Such configurations have been utilized in deriving
the kink-antikink scattering amplitudes in the bulk XXZ model (see
e.g. \cite{done1}) as well as in open XXZ chain with the most
general boundary conditions \cite{raj}, where the `duality'
$p^{\pm} \to -p^{\pm}$ cannot be implemented for the derivation of
the second eigenvalue of the boundary $S$-matrix. Notice that the
term depending on the boundary parameters (\ref{result1}) is
`double' compared to the diagonal case studied in \cite{jimbo,
done1}. Analogous phenomenon occurs in the open critical XXZ chain
\cite{doikounew, raj} and the sine--Gordon model \cite{GZ}. It is straightforward 
to see that in the diagonal limit we recover the results of \cite{jimbo, done1}. 
Also, in the isotropic limit $q \to 1$, $\Gamma_q(x) \to
\Gamma(x)$ and the trigonometric functions turn to rational, 
hence the generic rational reflection matrix for the open XXX spin chain is
easily recovered (see also \cite{macsh}).

It is finally convenient to rewrite the two eigenvalues in terms of
`renormalized' boundary parameters $\tilde p^{\pm}$ defined as:
\be \tilde p^{\pm} = p^{\pm} -{1\over 2} ~~~~~\mbox{mod}({i \pi
\over \eta}) \ee then the similarity between (\ref{result2}) and
the ratio of the `bare' eigenvalues (\ref{eigenv}) becomes
apparent. We have actually derived the physical boundary
$S$-matrix up to a gauge transformation; indeed the $S$-matrix of
the generic form (\ref{def}) may be reproduced by: \be {\mathrm K}(\lambda,\
\tilde p^+,\ \tilde p^-) = {\cal M}(\tilde p^+,\ \tilde p^-)\
\mbox{diag}\Big ({\mathrm k}_1(\lambda),\ {\mathrm k}_2(\lambda)
\Big )\ {\cal M}^{-1}(\tilde p^+,\ \tilde p^- ) \ee ${\cal M}$ is
defined in (\ref{gauge}). This concludes our derivation of the
general boundary $S$-matrix for the open XXZ chain in the
non-critical regime.
\\
\\
\noindent{\bf Acknowledgments:} This work was supported by INFN, 
Bologna section, through grant TO12.


\newpage 

\begin{thebibliography}{99}


\bibitem{chin} J. Cao, H.-Q Lin, K.-J. Shi and Y. Wang, Nucl. Phys. {\bf B663} (2003) 487.

\bibitem{nepo} R.I. Nepomechie, J. Stat. Phys. {\bf 111} (2003) 1363.

\bibitem{nemu} R. Murgan, R.I. Nepomechie and C. Shi, J. Stat. Mech. 0608 (2006) P006.


\bibitem{base} P. Baseilhac and K. Koizumi, {\it Exact spectrum of the XXZ open 
spin chain from the q-Onsager algebra representation theory}, hep-th/0703106.

\bibitem{galleas} W. Galleas,  {\it Functional relations from the 
Yang-Baxter algebra: Eigenvalues of the XXZ model with non-diagonal 
twisted and open boundary conditions}, arXiv:0708.0009.

\bibitem{jimbo}
M. Jimbo, R. Kedem, T. Kojima, H. Konno and T. Miwa,
Nucl. Phys. {\it B441} (1995) 437.

\bibitem{done1} A. Doikou, L. Mezincescu and R.I. Nepomechie, J. Phys. {\bf A31} (1998)
53.

\bibitem{FT} L.D. Faddeev and L.A. Takhtajan, J. Sov. Math. {\bf 24} (1984)241;\\
L.D. Faddeev and L.A. Takhtajan, Phys. Lett. {\bf 85A}
(1981) 375.

\bibitem{korepin}
V.E. Korepin, Theor. Math. Phys. {\it 76} (1980) 165;
V.E. Korepin, G. Izergin and N.M. Bogoliubov, {\it Quantum Inverse
Scattering Method, Correlation Functions and Algebraic Bethe Ansatz}
(Cambridge University Press, 1993).

\bibitem{AD}
N. Andrei and C. Destri, Nucl. Phys. {\it B231} (1984) 445.

\bibitem{FS}
P. Fendley and H. Saleur, Nucl. Phys. {\it B428} (1994) 681.

\bibitem{donebr} A. Doikou and R.I. Nepomechie, J. Phys, {\bf A32} (1999) 3663.

\bibitem{GZ}
S. Ghoshal and A. B. Zamolodchikov, Int. J. Mod. Phys. {\it A9} (1994)
3841; {\it A9} (1994) 4353.

\bibitem{doikounew} A. Doikou, {\it Generic boundary scattering in the open XXZ chain}, arXiv:0711.0716.

\bibitem{raj} R. Murgan, {\it  Boundary S matrix of an open XXZ spin chain with nondiagonal boundary terms}, arXiv:0711.1631.

\bibitem{ddvb} A. LeClair, G. Mussardo, H. Saleur and S. Skorik, Nucl. Phs. {\bf B453} (1995) 581;\\
C. Ahn, M. Bellacosa and F. Ravanini, Phys. Lett. {\bf B595} (2004) 537.

\bibitem{nepoh} C. Ahn and R.I. Nepomechie, Nucl. Phys. {\bf B676} (2004) 637;\\
C. Ahn, Z. Bajnok, R.I. Nepomechie, L. Palla and G. Takacs, Nucl. Phys. {\bf B714} (2005) 307.

\bibitem{ddv} A. Klumper, M.T. Batchelor and P.A. Pearce, J. Phys. {\bf A24} (1991) 3111;\\
C. Destri and  H.J. de Vega, Phys. Rev. Lett. {\bf 69} (1992) 2313.

\bibitem{DVGR} H. J. de Vega and A. Gonzalez--Ruiz, J. Phys. {\bf A27} (1994) 6129.

\bibitem{GMN} M.T. Grisaru, L. Mezincescu and R.I. Nepomechie, J. Phys. {\bf A28} (1995) 1027.

\bibitem{DMN}
A.  Doikou, L.  Mezincescu and R.I. Nepomechie, J. Phys. {\it A30}
(1997) L507.

\bibitem{GR}
G.  Gasper and M.  Rahman, {\it Basic Hypergeometric Series}
(Cambridge University Press, 1990).

\bibitem{macsh} N.J. MacKay and B.J. Short, Commun. Math. Phys. {\bf 233} (2003) 313;\\ Erratum-ibid. 245 (2004) 425.

\bibitem{sklyanin} E.K. Sklyanin, J. Phys. {\bf A21} (1988) 2375.

\bibitem{cherednik} I.V. Cherednik, Theor. Math. Phys. {\bf 61} (1984) 977.

\bibitem{doma} A. Doikou and P.P. Martin, J. Phys. {\bf A36} (2003) 2203.

\bibitem{annecy2} D. Arnaudon, J. Avan, N. Crampe, A. Doikou, L. Frappat and E. Ragoucy, J. Stat. Mech. 0408 (2004) P005.

\bibitem{doikouj} A. Doikou, J. Stat. Mech.  0605 (2006) P010.

\bibitem{doikous} A. Doikou, Nucl. Phys. {\bf B668} (2003) 447;\\
A. Doikou, Phys. Lett. {\bf A366} (2007) 556.

\bibitem{annecy} L. Frappat, R.I. Nepomechie and  E. Ragoucy,  J. Stat. Mech. 09 (2007) P09008

\bibitem{refst} W.-L. Yang, Y.-Z. Zhang, JHEP 04 (2007) 044.



\end{thebibliography}
\end{document}